\documentclass[aps,prl,superscriptaddress,twocolumn,showpacs,floatfix]{revtex4}

\usepackage{graphicx}

\def\ital{}

\begin{document} 

\title{
DNA double helices for single molecule electronics
}

\author{A. V. Malyshev}
\affiliation{
GISC, Departamento de F\'{i}sica de Materiales,
Universidad Complutense, Madrid, Spain\\
and A. F. Ioffe Physico-Technical Institute, 194021 St. Petersburg, Russia
}


\date{\today}

\begin{abstract}

The combination of self-assembly and electronic properties as well as its
true nanoscale dimensions make the DNA a promising candidate for the
building block of single molecule electronics. We argue that the intrinsic
double helix conformation of the DNA strands provides a possibility to drive
the electric current through the DNA by the perpendicular electric (gating)
field. The transistor effect in the poly(G)-poly(C) synthetic DNA is
demonstrated within a simple model approach. We put forward experimental
set-ups to observe the predicted effect and discuss possible device
applications of the DNA. In particular, we propose a design of the single
molecule analog of the Esaki diode.

\end{abstract}


\maketitle


The controversial question of charge transport in DNA molecules has been
attracting a great deal of attention recently (see
Refs.~\onlinecite{Ratner99,Porath04,Enders04} for an overview). The interest
in the DNA transport properties is at least two-fold: on the one hand, the
charge migration is believed to be important for the radiation damage
repair~\cite{Dandliker97} and, on the other, DNA double helices are
expected to be particularly useful for molecular
electronics~\cite{Enders04,Mertig99,Keren03,Cuniberti05}. While random base
sequences are relevant for biological samples, artificially
created periodic DNA molecules~\cite{Tippin97}, such as the poly(A)-poly(T)
or poly(G)-poly(C), are probably the best candidates for novel device
applications. The electrical transport through dry and wet DNA has
extensively been studied both theoretically and experimentally and a variety
of results has emerged: the DNA has been reported to demonstrate
proximity-induced superconducting~\cite{Kasumov01},
metallic~\cite{Okahata98,Fink99,Rakitin01,Legrand06},
semiconducting~\cite{Porath00,Yoo01,Hwang02,Xu04,Cohen05} and
insulating~\cite{Braun98,Storm01} behavior. Contact related effects, the
impact of the environment, and the DNA base pair sequence lead to such
diversity of results. According to both theory and experiment, the dry
poly(G)-poly(C) synthetic DNA is a semiconductor: theoretical {\em ab
initio} calculations predict a wide-band-gap semiconductor behavior (see,
e.g., Ref.~\onlinecite{Artacho03}) while experimental measurements reveal
about 2V voltage gap at low temperature~\cite{Porath00}.

Many effects useful for molecular device applications have been
reported: rectification, the Kondo effect, the Coulomb blockade, etc. (see
Ref.~\onlinecite{Cuniberti05} for a recent overview). In this contribution, it is
demonstrated for the first time that the intrinsic helix conformation of the
DNA strands determines transport properties of gated DNA molecules. In
particular, we show that the electric current through the double helix DNA
(in the base stacking direction) can be driven by the perpendicular
gating field. We put forward new experimental set-ups to reveal the
predicted effect and discuss possible applications of the DNA.
In particular, we propose a design of the single molecule analog of
the Esaki diode.


Two approaches are widely used to describe the DNA: {\em ab initio}
calculations~\cite{Artacho03,Pablo00,Barnet01,Starikov03,Wang04,Hubsch05,Starikov05,Mehrez05} and 
model-based Hamiltonians~\cite{Iguchi97,Jortner98,Cuniberti02,Roche03a,Roche03b,Unge03,Orsogna03,Iguchi03,Yamada04,Apalkov05b,Gutierrez05a,Kohler05,Yamada05,Klotsa05,Gutierrez06,Macia06}. 
The former can provide a detailed description but is currently limited to
relatively short molecules (typically of the order of 10 base pairs long).
The latter is much less detailed but allows for addressing systems of
realistic length. Model-based approach can play an important complementary
role because it grasps usually the underlying physics. Often, it yields
quite satisfactory quantitative results as well.

Here, we focus on qualitative properties of the DNA and use therefore the
effective Hamiltonian approach. A variety of models and parameter sets are
being discussed (see Refs.~\onlinecite{Klotsa05,Gutierrez06} and references
therein). In order to address intrinsic properties of the DNA, we do not
consider any environment or complex contact related effects and keep the
formalism as simple as possible. Hence, we adopt the {\em minimum}
tight-binding ladder model that accounts for the double-stranded structure
of the DNA. The ladder model was introduced in Ref.~\onlinecite{Iguchi97}
and has widely been used since then (see
Refs.~\onlinecite{Iguchi03,Yamada04,Gutierrez06} and references therein).
The Hamiltonian of the model reads: 
\begin{eqnarray}
	\nonumber
	\sum_{s,n} 
	\left(\,
	\varepsilon_{sn} |s\,n\rangle\langle s\,n|
	- t\,|s\,n\textstyle{+}1\rangle\langle s\,n|+h.c.
	-\tau\,|\textstyle{-}s\,n\rangle\langle s\,n|
	\,\right)
	\\
	\nonumber
	+ \sum_{s,k}
	\left(\,
	\varepsilon_{s\,M} |s\,k\rangle\langle s\,k|
	- t_M |s\,k\textstyle{+}1\rangle\langle s\,k|+h.c.
	\,\right)
	\\
	\nonumber
	- \sum_{s} 
	\left(\,
	\Gamma_{sl}\,|s\,0\rangle\langle s\,1|+
	\Gamma_{sr}\,|s\,N\textstyle{+}1\rangle\langle s\,N|+h.c.
	\,\right)
	\ ,
\end{eqnarray}
where the first term is the Hamiltonian of the $N$-base-pair DNA:
$\varepsilon_{sn}$ are on-site energies of base molecules
with index $n\in[1,N]$ labeling a pair and index $s=\pm 1$ labeling a
strand, $t$ and $\tau$ are inter-base hoppings parallel and perpendicular to
the base stacking direction, respectively. The second term describes
semi-infinite source ($k<1$) and drain ($k>N$) metallic leads with
$\varepsilon_{sM}=E_F$ and $t_M=4\,t$~\cite{Gutierrez06}, while the third is
the DNA-contact coupling term with
$\Gamma_{sl}=\Gamma_{sr}=t$~\cite{Roche03b,Gutierrez06}. $\langle sn|$ and
$|sn\rangle$ are bra and ket vectors of an electron at site $n$ of the
strand $s$.

Here, we extend the traditional ladder model (which neglects the helix
geometry of the strands) to the case when a molecule is subjected to the
perpendicular electric field and the helix conformation of the strands
becomes important. The B form of the DNA with the 10-base-pairs full-twist
period will be considered. Neglecting the difference between major and minor
grooves we set the on-site energies $\varepsilon_{sn}$ as follows:
\begin{equation}
	\varepsilon_{sn} = \varepsilon_{sn}^{(0)} + 
	e\,E_n\, s r\,\cos{\left(\frac{2\pi n}{10} + \varphi_0\right)}\ ,
	\label{e}
\end{equation}
where $\varepsilon_{sn}^{(0)}$ is site energy of the $sn$-th base
molecule at zero field, $E_n$ is the perpendicular gating field (for
simplicity, let it be homogeneous: $E_n=E_0$), and $r \sim 1$ nm is the
strand radius. Hereafter, we use the notation $V_g=2 E_0 r$ for the {\em
gate} voltage drop across the double helix. The phase $\varphi_0$ that
determines the orientation of the molecule with respect to the field is set
to $0$ from now on.

Equation~\ref{e} demonstrates that the perpendicular electric field results
in the harmonic modulation of the potential along the helical strands. The
modulation changes the electronic structure of the DNA and turns up to be
crucial for transport properties as we show below. Moreover, the amplitude
of the modulation can be controlled by the gating field, providing a
mechanism to alter the fundamental properties of the system.

\begin{figure}[b]
\includegraphics[clip,width=\columnwidth]{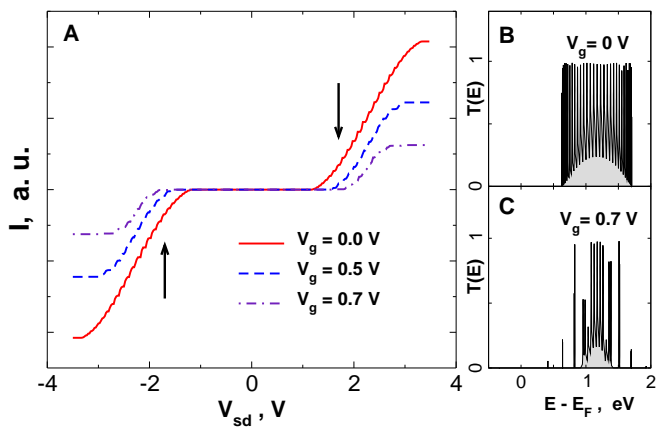}
\caption{
{\bf A} -- Current-voltage characteristics of the DNA at different gate
voltages $V_g$. {\bf B} and {\bf C} -- transmission coefficient for $V_g=0$
and $V_g=0.7V$. 
%
%
}
\label{I-Vsd}
\end{figure}

Throughout the paper, the electron transport through a 31-base-pairs
poly(G)-poly(C) DNA molecule is addressed with the following 
LUMO state on-site energies which are used as a starting point:
$\varepsilon_{-1n}^{(0)}=\varepsilon_G=1.14$~eV,
$\varepsilon_{+1n}^{(0)}=\varepsilon_C=-1.06$~eV~\cite{Mehrez05}. 
Hopping integrals $t$ and $\tau$ are not considered as bare tight-binding
parameters, rather as effective ones~\cite{Gutierrez06}, and are adjusted to
reproduce the voltage gap of about $2V$ which was observed in experiments on
the dry poly(G)-poly(C) DNA~\cite{Porath00,Cohen05}: $t=0.27 eV$, $\tau=0.25
eV$. These values are within reasonable parameter intervals~\cite{Macia06}.
Using the transmitting quantum boundary method (see
Refs.~\onlinecite{Lent90,Ting92} and references therein), we obtained the
transmission coefficient of the system,
$T(V_g,E)$, and calculated the current-voltage characteristics within the
scattering formalism~\cite{Ferry}:
\begin{eqnarray}
	I=\frac{2e}{h}\int T(V_g,E)\,\left(\,f_s(E,V_{sd})-f_d(E,V_{sd})\,\right)\,dE\ ,
	\nonumber
	\label{I}
\end{eqnarray}
where $f_{s,d}(E,V_{sd})=(1+e^{(E_F \pm e V_{sd}/2-E)/k T})^{-1}$ are
Fermi functions of source and drain contacts, $V_{sd}$ is the {\em
source-drain} voltage drop, and $E_F$ is the Fermi energy at
equilibrium taken to be in the middle of the DNA band gap, as for Au
contacts~\cite{Xu05}. The temperature $T$ is set to $4K$.


Figure \ref{I-Vsd} shows current-voltage characteristics and transmission
coefficients of the poly(G)-poly(C) DNA molecule at different gate
voltages. In all cases the system behaves as a semiconductor with the
voltage gap that varies with the gating field. Thus, within a range of
source-drain voltages (in the vicinity of those indicated by vertical arrows
in Fig.~\ref{I-Vsd}A), the system can be either conducting or insulating,
depending on the field.

\begin{figure}[t]
\includegraphics[clip,width=\columnwidth]{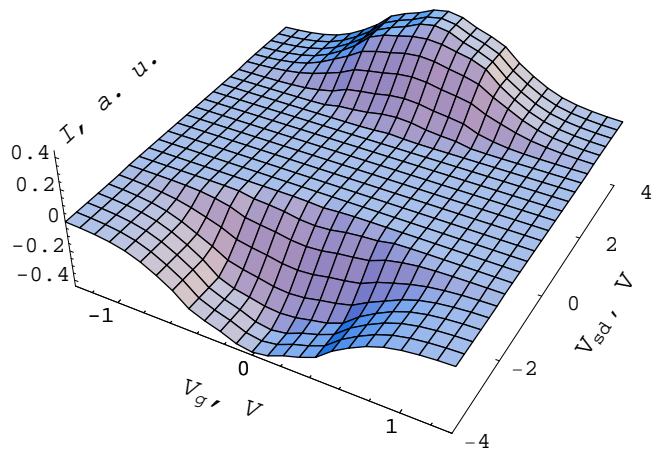}
\caption{
Current-voltage characteristics of the 
DNA molecule within a wide range of source-drain and gate voltages.
}
\label{I-V-3D}
\end{figure}

\begin{figure}[b]
\includegraphics[clip,width=\columnwidth]{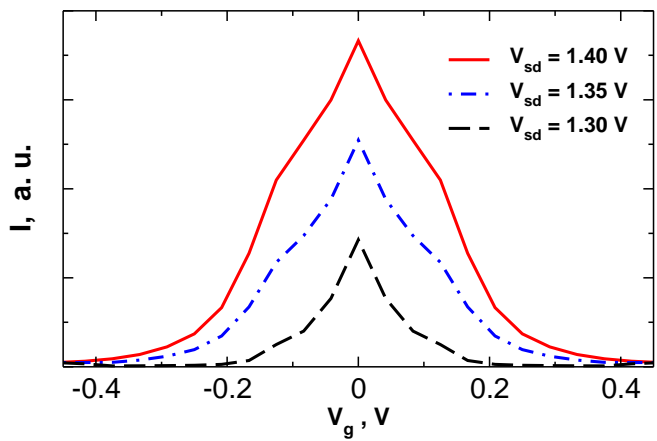}
\caption{
The dependence of the DNA source-drain current on the gate voltage $V_g$ at
different {\em fixed} source-drain voltages $V_{sd}$, indicated in the plot.
}
\label{I-Vg}
\end{figure}

Figure~\ref{I-V-3D} demonstrates current-voltage
characteristics within a wide range of source-drain and gate voltages.
Semiconducting behavior can be observed for $|V_g| \le 0.7V$ as well as
strong gating effect for $|V_{sd}| \ge 1.2V$. To illustrate the gating
effect, we plot in Fig.~\ref{I-Vg} dependencies of the
current on the gate voltage drop for several {\em fixed} values of the
source-drain voltage $V_{sd}$. For all values of $V_{sd}$, a typical
hat-like $I$--$V_g$ characteristic is observed. Strong dependence of the
source-drain current on the gate voltage suggests the usage of the gated
double helix DNA as a field-effect transistor.

\begin{figure}[t]
\includegraphics[clip,width=\columnwidth]{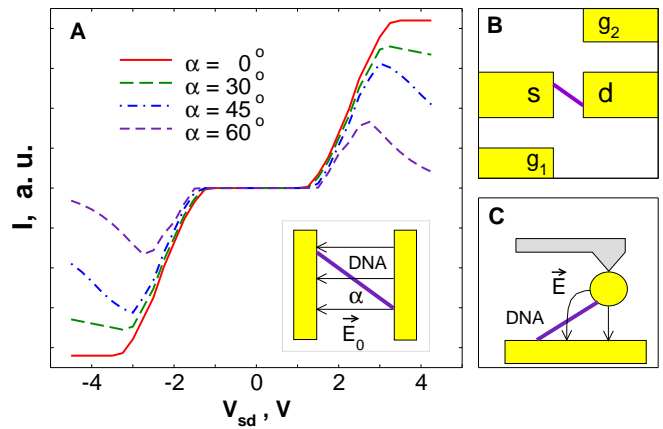}
\caption{
{\bf A} -- Non-monotonous current-voltage characteristics of the single
molecule analog of the Esaki diode for different angles $\alpha$ between the
tilted molecule and the inter-contact electric field; inset --- a scheme of
the device. ({\bf B},{\bf C}) -- proposed experimental set-ups to observe
the predicted non-monotonous $I$--$V$ curves. Panel {\bf B} illustrates the
four contact configuration that is put forward for {\em asymmetric}
electrostatic trapping of a DNA molecule: applying appropriate biases to all
four contacts a tilted electrostatic trapping field can be induced between
the electrodes ``s'' and ``d''. A polarized molecule aligns with the field
and gets trapped between contacts ``s'' and ``d'' being tilted with respect
to them. Electrodes ``s'' and ``d'' can then be used for measurements,
resulting in the experimental configuration presented in the inset of panel
{\it A}. 
%
%
See the text for comments on panel {\bf C}.
}
\label{I-Vsd-angle}
\end{figure}

To perform conductance measurements, a linear DNA molecule is usually
trapped between two contacts (see, e.g.,
Refs.~\onlinecite{Porath00,Cohen05}). If the trapped molecule is not aligned
with the inter-contact electric field then there exists the component of the
field perpendicular to the molecule axis (see the inset of
Fig.~\ref{I-Vsd-angle}A), which produces the gating effect. For such tilted
molecule of length $L$, the gate voltage drop depends on the source-drain
voltage:
\begin{eqnarray}
	V_g=\frac{2r}{L}\,\tan{(\alpha)}\,V_{sd}\ .
	\nonumber
	\label{Vg}
\end{eqnarray}
Thus, on the one hand, the current tends to increase with the source-drain
voltage, while on the other, higher source-drain voltage leads to stronger
gating, which tends to reduce the current. The interplay of the two
opposite contributions can lead to new physical effects.

Current-voltage characteristics of a tilted DNA for different angles
$\alpha$ between the molecule and the homogeneous inter-electrode field are
presented in Fig.~\ref{I-Vsd-angle}A.  The $I$-$V$ curves are non-monotonous
and have a region with the negative differential resistance, similar to
those of the tunneling diode, which suggests that the proposed device is a
single molecule analog of the Esaki diode.

We performed similar calculations for the poly(A)-poly(T) DNA, other
parameter sets (e.g., ionization energies~\cite{Sugiyama96,Voityuk01},
HOMO/LUMO energies from Ref.~\onlinecite{Mehrez05}), gating electric field
profiles, temperatures, and within the framework of the dangling backbone
ladder model~\cite{Klotsa05} which accounts for both DNA bases and
backbones. The gating effect was found to be generic. We note that the ratio
of the DNA length to its full-twist period, the phase $\varphi_0$, and other
factors that change the symmetry of the system can result in some
qualitative changes. Nevertheless, the strong dependence of the source-drain
current on the perpendicular field remains intact.

The underlying gating effect is the direct consequence of the helix geometry
of the DNA strands only: the modulation of the strand potential by the
gating field (see Eq.~\ref{e}) modifies the energy spectrum and reconstructs
transmission bands. At non-zero gating field each band splits into several
mini-bands that are degrading as the field increases; outer mini-bands
degrade faster, which leads to the effective increase of transmission and
voltage gaps (see Fig.~\ref{I-Vsd}). This mechanism should be taken into
account for correct interpretation of experiments with electrostatically
coupled gate.


With a view to observe the predicted effect, the following experimental
set-ups can be proposed. In Ref.~\onlinecite{Porath00} a poly(G)-poly(C) DNA
molecule was deposited between two contacts by the electrostatic
trapping~\cite{Bezryadin97}. If a molecule can deliberately be put at a
sufficiently large angle to the contacts then the component of the
inter-contact field perpendicular to the molecule would produce the sought
gating effect as we argue above. During the electrostatic trapping, a DNA
molecule becomes polarized by the trapping field and is attracted to the
volume where the field is maximum (i.e., to the inter-electrode region). The
polarized molecule tends to align with the field, thus, asymmetric
deposition can be achieved by creating an asymmetric trapping field between
the electrodes (see caption of Fig.~\ref{I-Vsd-angle}{B} for details).

Another experimental set-up was discussed in Ref.~\onlinecite{Cohen05} where
a DNA molecule was trapped between a substrate and a golden nano-particle
suspended from the metal-coated tip of an AFM microscope. It should be
possible to keep the DNA molecule tilted with respect to the substrate by
appropriate displacements of the tip, in which case there would exist a
component of the tip-substrate electric field perpendicular to the DNA base
stacking direction (see Fig.~\ref{I-Vsd-angle}{C}). We note that the rise of
the current that was observed during the retraction of the tip from the
surface at constant tip voltage~\cite{Cohen05} can be explained by the
gating effect, which confirms indirectly our prediction. It would therefore
be desirable to measure the current-voltage characteristic of a constantly
tilted molecule, which is expected to be non-monotonous.


In summary, we have demonstrated for the first time that the intrinsic helix
conformation of DNA strands has strong impact on transport properties of the
molecule. We consider the periodic DNA and show that the electric current
through it (in the base stacking direction) can be driven by the
perpendicular electric field, suggesting such applications as the field
effect transistor.

We propose also a new molecular device: the periodic DNA trapped between two
contacts at an appropriate angle to them (at about $45^\circ$). The
current-voltage characteristic of such device is non-monotonous and has a
region of the negative differential resistance, analogous to that of the
Esaki tunneling diode.

To conclude, the predicted gating effect opens a possibility to use the DNA
for various novel molecular devices. The same argumentation may also apply
to G4-DNA and proteins many of which have the $\alpha$-helix
conformation~\cite{Branden}.


The author is grateful to F. Dom\'{i}nguez-Adame, E. Maci\'{a}, R.
Gutierrez, and V. Malyshev for fruitful discussions, VM's constant
encouragement throughout the study and comments on the manuscript are
appreciated. This study was supported by MEC under projects 
Ram\'{o}n y Cajal and MOSAICO.

\end{document}